\documentclass[review]{elsarticle}

\usepackage{blindtext, graphicx}
\usepackage[utf8]{inputenc}
\usepackage[english]{babel}
\usepackage{listings}
\usepackage{amssymb}
\usepackage[breaklinks=true]{hyperref}
\usepackage{booktabs}
\usepackage{multirow}
\usepackage{breqn}
\usepackage{scrextend}
\usepackage{soul,color}
\usepackage{ulem}

\usepackage{xcolor}

\usepackage{graphicx}

\usepackage[noend,linesnumbered,ruled,vlined]{algorithm2e}
\usepackage[font=footnotesize,labelfont=bf]{caption}
\usepackage[font=footnotesize,labelfont=bf]{subcaption}

\begin{document}

\begin{frontmatter}
\title{Modularity-based approach for tracking communities\\ in dynamic social networks}

\author[1]{Michele Mazza}
\ead{michele.mazza@iit.cnr.it}

\author[1]{Guglielmo Cola\corref{cor1}}
\ead{guglielmo.cola@iit.cnr.it}

\author[1]{Maurizio Tesconi}
\ead{maurizio.tesconi@iit.cnr.it}

\address[1]{Institute of Informatics and Telematics, National Research Council, Via G. Moruzzi, 1, 56124, Pisa, Italy}
\cortext[cor1]{Corresponding author}

\begin{abstract}

\vspace{10pt} 
\noindent\colorbox{yellow}{%
  \parbox{1.0\textwidth}{%
    \color{red}
    \textbf{Note}: This is a preprint version. The final, peer-reviewed version has been published in Knowledge-Based Systems (Elsevier). Please refer to and cite the published version: \url{https://doi.org/10.1016/j.knosys.2023.111067}.
  }%
}
\vspace{10pt} 

Community detection is a crucial task to unravel the intricate dynamics of online social networks. The emergence of these networks has dramatically increased the volume and speed of interactions among users, presenting researchers with unprecedented opportunities to explore and analyze the underlying structure of social communities. Despite a growing interest in tracking the evolution of groups of users in real-world social networks, the predominant focus of community detection efforts has been on communities within static networks. In this paper, we introduce a novel framework for tracking communities over time in a dynamic network, where a series of significant events is identified for each community. Our framework adopts a modularity-based strategy and does not require a predefined threshold, leading to a more accurate and robust tracking of dynamic communities. We validated the efficacy of our framework through extensive experiments on synthetic networks featuring embedded events. The results indicate that our framework can outperform the state-of-the-art methods. Furthermore, we utilized the proposed approach on a Twitter network comprising over 60,000 users and 5 million tweets throughout 2020, showcasing its potential in identifying dynamic communities in real-world scenarios. The proposed framework can be applied to different social networks and provides a valuable tool to gain deeper insights into the evolution of communities in dynamic social networks.
\end{abstract}


\end{frontmatter}

\section{Introduction}
With the rapid proliferation and evolution of various social networking systems, such as online social networks, mobile networks~\cite{wu2009}, and collaboration networks~\cite{newman2001}, social network analysis has emerged as a critical research topic. In this field, community detection plays a key role in uncovering meaningful group structures within networks, offering insights into the underlying social dynamics. Hence, various community detection algorithms have been proposed, considering factors such as individual cooperative or hostile relationships~\cite{xia22}, and the utilization of local information to identify hidden communities~\cite{luo19}.
Despite the significant attention that community detection has received in recent literature, most approaches have focused on static networks, overlooking the temporal properties inherent in real-world networks. Traditional methods rely on aggregation to build a static network that represents all the interactions over a specified period~\cite{wasserman1994}. Inevitably, such representation is unable to capture the evolving nature of communities influenced by time-sensitive events. 
For example, in a Twitter retweet network~\cite{CINELLI2022113819, zola2020} the evolution of community structures can be driven by factors like new users starting to retweet a specific account or existing community members ceasing their interactions. Treating these types of networks as static can lead to invalid associations and an oversimplified representation of communities. A user may show similarities with a community at a certain time and later move towards a different community: this user would represent the only connection between the two communities, and omitting the temporal properties may lead to the aggregation of these communities as a single large community. In addition, time can also affect content semantically. For example, the meaning of the hashtag \#MeToo changed dramatically with the emergence of the \#MeToo social movement. 
Thus, community detection in such networks is crucial to enable a deeper understanding of the underlying dynamics.

A dynamic network can be represented as a time series of static networks called snapshots~\cite{wolf2006}. Each snapshot corresponds to the interactions aggregated over a defined period, such as a week or an hour. An intuitive method to detect communities in a dynamic network partitioned into snapshots consists in employing well-studied static community detection algorithms on each snapshot. Next, dynamic communities can be tracked by identifying the events that shape the evolution of a community over time~\cite{greene2010tracking, takaffoli2011, brodka2013ged}. This step involves matching the communities found at different snapshots through an algorithm, usually based on the similarity of community members. Therefore, an arbitrary threshold is required to determine if two or more communities found at different snapshots are indeed the result of the evolution of the same community over time. The main drawback is that the final result may change drastically depending on the selected threshold: a high similarity threshold leads to less tolerance for fluctuating members, whereas setting a too-low similarity threshold leads to the merge of dissimilar communities. The scarcity of threshold-free frameworks negatively affects their real-world implementation, which could instead prove useful in countering threats that plague online social platforms such as disinformation~\cite{MARTIN2022109265, QURESHI2022109649} and information manipulation~\cite{SUCHACKA2020105875, LI2022110038}.

In this paper, we introduce a threshold-free framework crafted to track the evolution and structure of communities across multiple snapshots of a dynamic network. Given the communities detected across all snapshots, we represent the entire network as a similarity network, where the nodes are the static communities found and the weight of the edges corresponds to their similarity. By applying \textit{Local Modularity Optimization}~\cite{blondel2008}, we reveal the groups of nodes having high modularity and similarity. In this manner, we turn the community matching task into a modularity optimization problem where dynamic communities are found by optimizing modularity locally on all nodes. The node groups found embody the evolution of a community over time. Finally, each group is disentangled to reconstruct the temporal evolution of the represented dynamic community.

Notably, by constructing a single network encompassing all static communities across snapshots, our approach avoids sequential matching, which struggles to track intermittent communities that may not be present in consecutive snapshots. Through local modularity optimization, we derive groups of communities with robust interconnections and marked similarities, at the same time preserving a desirable degree of flexibility between them. Importantly, our threshold-free design guarantees consistent and trustworthy outcomes, promoting the application of our framework in real-world scenarios.

For the evaluation of our framework, we used four synthetic dynamic networks~\cite{greene2010tracking} embedded with distinct communities and events. Results indicate that, in most cases, our approach performs better than the state-of-the-art community tracking methods. In the second part of our evaluation, we show how the proposed approach can be applied to a real-world co-hashtag Twitter network~\cite{mazza2022} to extract relevant information about dynamic communities.  

The main contributions of our work can be summarized as follows:
\begin{itemize}
    \item We propose a novel framework for tracking the evolution and structure of communities in dynamic networks, leveraging local modularity optimization. The ability to track communities in dynamic networks is of paramount importance to fully understand the underlying social dynamics of real-world networks, like online social networks.

    \item Our framework offers a significant advantage over existing methods by eliminating the need for setting a specific threshold, leading to more accurate and robust tracking of dynamic communities.

    \item The evaluation of our novel framework on four synthetic datasets demonstrates its superior performance compared to the state of the art.

    \item We showcase the practical application of our framework on a Twitter dataset generated by real malicious accounts, illustrating its potential to extract relevant and suggestive information about dynamic communities.
\end{itemize}

The subsequent section offers a brief overview of existing research in dynamic community tracking. In Section~\ref{sec:foundations}, we provide foundational concepts pertinent to dynamic networks. Section~\ref{sec:proposed} delves into a comprehensive description of our proposed framework. Section~\ref{sec:evaluation} encompasses both the evaluation and comparison of our framework against other methodologies and the presentation of results derived from a real-world network. Finally, Section~\ref{sec:conclusions} wraps up the paper with a concise summary and potential avenues for future research.

\section{Related work}
\label{sec:related}
The inherent dynamic nature of most social networks~\cite{newman2003social} has spurred significant research interest in studying their evolution in recent years. In this section, we briefly review works that, similar to our approach, adapt existing static community detection algorithms to capture the temporal dynamics of social networks~\cite{KADKHODAMOHAMMADMOSAFERI2020113221}. This approach is particularly suitable for social networks with highly dynamic community structures~\cite{takaffoli2010framework}.
While here we provide an overview of the key contributions from each work, a detailed examination of selected studies is provided in Section~\ref{sec:evaluation}.

One pioneering work that utilized static network snapshots to track community evolution over time is presented in~\cite{hopcroft2004tracking}. The authors proposed a method based on agglomerative hierarchical clustering to identify and track stable clusters over time.
In~\cite{palla2007quantifying}, the clique percolation method was extended to monitor events in the evolution of dynamic networks.
They built joint graphs for pairs of consecutive snapshots and then matched the clique-based communities obtained using an autocorrelation function to find overlap between two states of a community. 
The work presented in~\cite{asur2009event} introduced a strategy, implemented as bit operations, to identify events between communities found in two consecutive snapshots.
Another efficient approach to identify and track dynamic communities across multiple snapshots is described in~\cite{greene2010tracking}. The authors employed a community matching strategy based on weighted bipartite matching. 
A framework proposed by~\cite{takaffoli2010framework} offers an event-based approach for detecting transitions between communities in consecutive snapshots. In a later work from the same authors~\cite{takaffoli2011}, the event definition formula has been improved to track community transitions throughout the observation time, no longer restricting it to consecutive snapshots.
The Group Evolution Discovery (GED) framework, introduced by~\cite{brodka2013ged}, takes into account not only the similarity of community members but also the positions and importance of nodes within the community. This approach facilitates matching communities and tracking their evolution across consecutive snapshots. 
More recent studies~\cite{tajeuna2015tracking, tajeuna2016tracking} proposed a method to detect and model the evolution of a community using a novel similarity measure termed ``mutual transition''. In~\cite{KADKHODAMOHAMMADMOSAFERI2020113221}, a method for the Identification of Community Evolution by Mapping (ICEM) has been presented. 
In their approach, the evolution of a community is tracked by representing community members within a hash map.

We provide a more detailed description and analysis of~\cite{greene2010tracking, takaffoli2011, takaffoli2010framework, brodka2013ged, tajeuna2015tracking, tajeuna2016tracking, KADKHODAMOHAMMADMOSAFERI2020113221} in Section~\ref{sec:evaluation}, where we compare the strengths and weaknesses of these methods with respect to our novel framework.

\section{Foundational concepts on dynamic networks}
\label{sec:foundations}

Before delving into our framework, let us first introduce foundational concepts pertinent to dynamic social networks.

\subsection{Dynamic communities}
\begin{figure}
  \centering
  \includegraphics[width=0.5\linewidth]{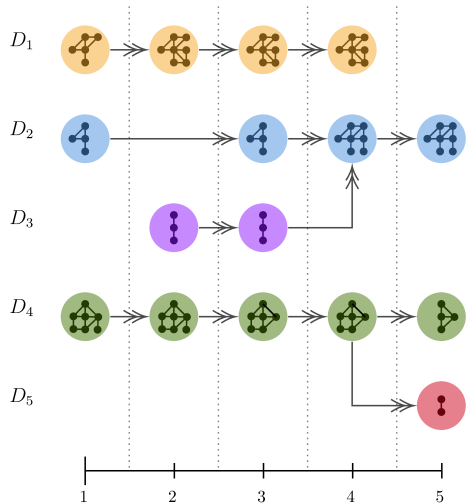}
  \caption{Example of five dynamic communities tracked over five snapshots, featuring growth, contraction, merging, splitting, and death events.}
  \label{fig:dynamic_schema}
\end{figure}

A dynamic social network can be represented as a series of $n$ graphs $\mathbb{G} = \{G_1, G_2, ..., G_n\}$, where $G_i$ represents the graph at the $i^{th}$ snapshot, encompassing only the nodes and edges present at that particular time. Next, we define as $\mathbb{C}^i = \{C^i_1, C^i_2, ..., C^i_{k_i}\}$ the set of $k_i$ \textit{static} communities found at the $i^{th}$ snapshot. Our objective is to find a set of $m$ \textit{dynamic} communities $\mathbb{D} = \{D_1, D_2, ..., D_m\}$ that occur in one or more snapshots. Each dynamic community $D_j$ is represented by a timeline of its constituent static communities ordered in time. Figure~\ref{fig:dynamic_schema} shows an example of timelines for five dynamic communities. 
When dynamic communities, such as $D_2$ and $D_3$ or $D_4$ and $D_5$, have interconnected static communities within them, we can either treat these dynamic communities independently or merge them into a single entity that encompasses all their constituent static communities. In the latter case, they become $D_{(2,3)}$ and $D_{(4,5)}$.

\subsection{Critical events}
In order to track communities and their structural changes over time, we need to define structural events to describe the evolutionary behavior of dynamic social networks.
In~\cite{palla2007quantifying} and~\cite{asur2009event}, some types of events are proposed, but they do not cover all the possible ways in which a community can evolve. This paper uses the same models of critical events as proposed in~\cite{greene2010tracking}:
\textit{Growth} (a community gains new members), 
\textit{Contraction} (a community loses some members), 
\textit{Merging} (two or more communities merge into a new one), 
\textit{Splitting} (a community is split into two or more new ones), 
\textit{Birth} (a new community appears), 
\textit{Death} (a community disappears).

Examples of these events are illustrated in Figure~\ref{fig:dynamic_schema}:

\begin{itemize}
    \item The community in $D_1$ \textit{grows} at the $2^{nd}$ snapshot.
    \item The community in $D_4$ \textit{contracts} at the $3^{rd}$ snapshot.
    \item The community in $D_3$ \textit{merges} with the community in $D_2$ at the $4^{th}$ snapshot.
    \item The community in $D_4$ \textit{splits} into two communities at the $5^{th}$ snapshot, one still in $D_4$ and the other in $D_5$.
    \item The communities in $D_1$, $D_2$, and $D_4$ were \textit{born} at the $1^{st}$ snapshot, whereas the community in $D_3$ emerged at the $2^{nd}$ snapshot and the community in  $D_5$ was formed at the $5^{th}$ snapshot.
    \item The community in $D_1$ experienced a \textit{death} event at the $4^{th}$ snapshot.
\end{itemize}

It is worth noting that a community may not manifest at every snapshot. Considering the example in Figure~\ref{fig:dynamic_schema}, the community in $D_2$ is observed at the $1^{st}$ snapshot and then again at the $3^{rd}$ snapshot. This ``intermittence'' can be caused by the behavior of community members or depend on the duration granularity of each snapshot.

\subsection{Similarity}

When tracking dynamic communities and the evolutionary events of their constituent communities, the key to finding relationships between communities from different snapshots is similarity. Most existing methods require setting a similarity threshold: communities are deemed related only if their similarity exceeds the threshold. In evolving dynamic social networks, unstable communities might undergo member shifts, losing original members and gaining new ones over time. 
Hence, setting the similarity threshold too high fails to identify these evolving communities, while a threshold set too low may erroneously link unrelated communities. As detailed in the following section, our framework utilizes similarity without requiring a predefined threshold.

\section{Proposed framework to track dynamic communities}
\label{sec:proposed}

\begin{figure*}[ht]
  \includegraphics[width=\textwidth]{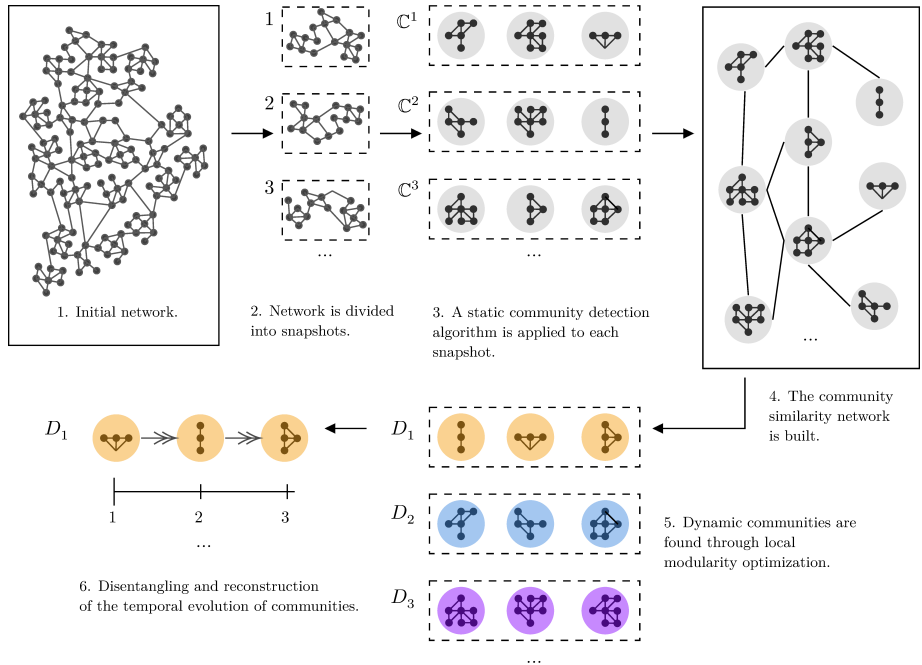}
  \caption{Workflow of the proposed approach. Starting from a network, we divide it into snapshots. Then, we apply a static community detection algorithm to each snapshot. We construct the community similarity network using the communities found at the previous step. Dynamic communities are subsequently identified using local modularity optimization. Finally, for each dynamic community, we reorder its static communities based on their originating snapshots to highlight critical events.
  }
  \label{fig:t_window}
\end{figure*}

Figure~\ref{fig:t_window} provides an outline of our framework. We initiate the process by segmenting the initial network into snapshots.
Next, we apply a static community detection method to each snapshot. Subsequently, we construct the community similarity network, where each static community is represented as a node.
In this phase, we disregard the origin of the static communities, enabling matches between communities from non-consecutive snapshots.
We then identify dynamic communities by optimizing modularity locally across all nodes. Lastly, for each dynamic community, we reorder its constituent communities to reconstruct its temporal evolution and identify the events that occurred across snapshots.

\subsection{Community similarity network}
In our framework, the matching step relies on a weighted undirected community similarity network $G_c=(V_c, E_c, W_c)$ where the vertex set (nodes) $V_c$ encompasses the static communities found at each snapshot, the edge set $E_c$ consists of links connecting these communities, and $W_c$ assigns weights to these edges based on the similarity values of the connected communities. Notably, an edge is established between two communities only if they originate from distinct snapshots and share a non-zero similarity.

We compute similarity as the \textit{overlap coefficient} between two communities. Let $\mathbb{C}^i$ and $\mathbb{C}^j$ denote the sets of communities found at the snapshots $i$ and $j$, where $i \neq j$. Further, let $\alpha  \in \mathbb{C}^i$ and $\beta \in \mathbb{C}^j$. The overlap coefficient between two communities, $\alpha$ and $\beta$, is calculated as follows:
\begin{equation}
overlap(\alpha, \beta) = \frac{|\alpha \cap \beta|}{min(|\alpha|,|\beta|)}
\end{equation}
where $|\alpha|$ and $|\beta|$ are the number of members, respectively, in $\alpha$ and $\beta$, while $|\alpha \cap \beta|$ is the number of shared members. The proposed framework is not tied to any specific similarity metric. For example, one could employ the Jaccard similarity, as shown in~\cite{greene2010tracking}.

\subsection{Local modularity optimization}

After building the similarity network, we identify groups of similar nodes by optimizing modularity locally on all nodes. 
Modularity is a leading metric for evaluating community structures~\cite{chen2013measuring}. Scaled between $[-0.5,1]$, it quantifies the density of intra-community links compared to inter-community links. By optimizing modularity, we seek a community assignment for each node in the network such that modularity $M$ is maximized using the function defined by:
\begin{equation}
    M={\frac {1}{2m}}\sum \limits _{xy}{\bigg [}A_{xy}-{\frac {k_{x}k_{y}}{2m}}{\bigg ]}\delta (c_{x},c_{y})
\end{equation}
where $A$ is the adjacency matrix with $A_{xy}$ representing the weight of the edge between the vertices $x$ and $y$, 
$k_x$ and $k_y$ are the sum of the weights of the edges attached to vertices $x$ and $y$, respectively,
$c_x$ and $c_y$ are the communities to which vertices $x$ and $y$ belong, $\delta$ is the Kronecker delta function ($\delta(c_x,c_y)$ returns 1 if $c_x=c_y$ and 0 otherwise), and $m$ is the sum of the weights of all edges in the graph. In modularity optimization, edge weights, which represent the connection strength, guide the optimization process. Initially, each vertex in the network is assigned to its own community. Then, each vertex $x$ is removed from its community and moved to the community of each neighbor vertex $y$. When a vertex is assigned to a new community, the modularity increase is calculated:
\begin{equation} 
\begin{gathered}
\Delta M={\bigg [}{\frac {\Sigma _{in}+2k_{x,in}}{2m}}-{\bigg (}{\frac {\Sigma _{tot}+k_{x}}{2m}}{\bigg )}^{2}{\bigg ]}-\\{\bigg [}{\frac {\Sigma _{in}}{2m}}-{\bigg (}{\frac {\Sigma _{tot}}{2m}}{\bigg )}^{2}-{\bigg (}{\frac {k_{x}}{2m}}{\bigg )}^{2}{\bigg ]}
\end{gathered}
\end{equation}
where $\Sigma _{in}$ is the sum of all the weights of the links within the community to which $x$ is moving, 
$\Sigma _{tot}$ is the sum of all the weights of the links to nodes in the community to which $x$ is moving,
$k_x$ is the weighted degree of $x$, $k_{x,in}$ is the sum of the weights of the edges between $x$ and the other nodes in the community to which $x$ is moving, and $m$ is the sum of the weights of all edges in the network. Then, once $\Delta M$ is calculated for all communities connected to $x$, the community that resulted in the most significant increase in modularity encompasses $x$. If there is no increase in modularity, vertex $x$ stays in its initial community. The process described above is applied to all vertices until there is no further increase in modularity. While our approach aligns with the initial phase of the Louvain algorithm~\cite{blondel2008}, the latter then transitions to a second phase, building a new network where the vertices represent the communities found at the previous phase. Next, it applies the first phase to this new network, further optimizing modularity. A complete run of both phases is called a pass. Such passes are carried out repeatedly until a maximum of modularity is achieved. As our goal is to aggregate vertices with high similarity rather than maximizing modularity, we only employ the first step. Furthermore, because of the intrinsic properties of the community similarity network, using Louvain's complete procedure might yield communities with low granularity, potentially leading to clusters where certain members are poorly related to each other. Therefore, by solely applying local modularity optimization to the vertices of the community similarity network, we obtain clusters with high modularity and thus composed of similar communities. These clusters correspond to the dynamic communities featured in the dynamic network.

\subsection{Identify dynamic communities over time}

The pseudocode to identify and track dynamic communities over time is presented in Algorithm~\ref{alg:tracking}. First, the community similarity network $G_c$ is initialized as an empty graph (Line 1). For each community within all snapshots, all subsequent snapshots are traversed to find communities with a similarity value greater than 0 (lines 2--7). When two communities exhibit a positive similarity, they are introduced as vertices in $G_c$, connected by an edge weighted by this similarity value (lines 8--9). 
Subsequently, a community set is initialized, where each node in $G_c$ is assigned to its own distinct community (line 10). 
Local modularity optimization is then applied, evaluating each vertex to determine if its movement to another community would bolster network modularity (lines 11--20). The final result is a collection of sets of communities, each of these sets being an identified dynamic community. By reordering the communities in each set based on time, we can uncover the events that shaped the evolution of the dynamic community.

\begin{algorithm}
    \SetAlgoLined
    \SetKwInOut{Input}{Input}
    \SetKwInOut{Output}{Output}
    \Input{S (List of sets of static communities contained at each snapshot)}
    \Output{O (Collection of sets, each set being an identified dynamic community)}

    $G_c\gets \{\}$ \tcp*{initialize $G_c$ with an empty network}
    \tcc{compare each community with all communities belonging to subsequent sets (snapshots)}
    \For{$\mathbb{C}^i \in S$}{
        \For{$\alpha \in \mathbb{C}^i$}{
            $S^j \gets$ all subsequent snapshots of $\mathbb{C}^i$\;
            \For{$\mathbb{C}^j \in S^j$}{
                \For{$\beta \in \mathbb{C}^j$}{
                     $sim = overlap(\alpha,\beta)$\;
                     \uIf{$sim > 0$}{
                        add an edge in $G_c$ between $\alpha$ and $\beta$, with weight $sim$\;
                    }
                }
            }
        }
    }
    \tcc{optimize local modularity of $G_c$}
    $O\gets \{\{v\} | v \in G_c(V) \}$\tcp*{initialize each node of $G_c$ with its own community}
    \While{one or more nodes are moved to a different community}{
        \For{$x \in G_c(V)$}{
            $best_q \gets -\infty$\;
            $best_c \gets$ community of $x$\;
            \For{all neighboring nodes $y$ of $x$}{
                $gain_q \gets \Delta M$ moving $x$ to the community of $y$\;
                \uIf{$best_q \le gain_q$}{
                        $best_q \gets gain_q$\;
                        $best_c \gets$ community of $y$\;
                    }
            }
            $O \gets$ place $x$ in $best_c$\;
        }
    }
return $O$\;
\caption{Tracking dynamic communities}
\label{alg:tracking}
\end{algorithm}

\subsection{Events reconstruction}
Once the dynamic communities have been found from the community similarity network, we proceed with the identification of the events that describe their evolutionary behavior. Given a dynamic community $D$, which is a set of static communities found across snapshots, let us define more formally each critical event as follows:
\begin{itemize}
    \item Growth -- A community $C^i \in D$ grows if there is a community $C^j \in D$ at a prior snapshot ($i > j$) that shares members with $C^i$ and is smaller in size
    \begin{equation*}
    \begin{gathered}
        growth(C^i) = 1 \qquad if \\
        \exists \ C^j \in D: \quad i>j, \quad \lvert C^i \cap C^j \rvert > 0, \quad \lvert C^i \rvert > \lvert C^j \rvert~.
    \end{gathered}
    \end{equation*}

    \item Contraction -- A community $C^i \in D$ contracts if there is a community $C^j \in D$ at a prior snapshot ($i > j$) that shares members with $C^i$ and is larger in size.
    \begin{equation*}
    \begin{gathered}
        contract(C^i) = 1 \qquad if \\
        \exists \ C^j \in D: \quad i>j, \quad \lvert C^i \cap C^j \rvert > 0, \quad \lvert C^i \rvert < \lvert C^j \rvert~.
    \end{gathered}
    \end{equation*}

    \item Merging -- A community $C^i \in D$ results from a merge if there is a set of communities $C^j_{set} \in D$ from a prior snapshot ($i > j$) such that each community in $C^j_{set}$ shares members with $C^i$
    \begin{equation*}
    \begin{gathered}
        merge(C^i) = 1 \qquad if \\
        \exists \ C^j_{set} = \{C^j_1, C^j_2, ..., C^j_n\}: \quad i>j, \\
        \forall \ C^j \in C^j_{set}: \quad \lvert C^i \cap C^j \rvert > 0~.
    \end{gathered}
    \end{equation*}

    \item Split -- A community $C^i \in D$ results from a split if it is part of a set of communities $C^i_{set} \in D$ at snapshot $i$ and there is a community $C^j \in D$ from a prior snapshot ($i > j$) such that each community in $C^i_{set}$ shares members with $C^j$
    \begin{equation*}
    \begin{gathered}
        split(C^i) = 1 \qquad if \\
        \exists \ C^i_{set} = \{C^i_1, C^i_2, ..., C^i_n\}: \quad i>j, \\
        \forall \ C^i \in C^i_{set}: \quad \lvert C^i \cap C^j \rvert > 0~.
    \end{gathered}
    \end{equation*}

    \item Birth -- A community $C^i \in D$ is born if there is no community $C^j \in D$ from any prior snapshot ($i > j$) that shares members with $C^i$
    \begin{equation*}
    \begin{gathered}
        birth(C^i) = 1 \qquad if \\
        \nexists \ C^j \in D: \quad i>j, \quad \lvert C^i \cap C^j \rvert > 0~.
    \end{gathered}
    \end{equation*}

    \item Death -- A community $C^i \in D$ dies if there is no community $C^j \in D$ from any subsequent snapshot ($j > i$) that shares members with $C^i$
    \begin{equation*}
    \begin{gathered}
        death(C^i) = 1 \qquad if \\
        \nexists \ C^j \in D: \quad j>i, \quad \lvert C^i \cap C^j \rvert > 0~.
    \end{gathered}
    \end{equation*}
\end{itemize}

\section{Experiments and results}
\label{sec:evaluation}

In this section, we detail the experiments conducted to compare our framework with state-of-the-art methods. Additionally, we present an example utilizing real-world data to demonstrate our framework's capability in extracting valuable insights from social interactions.

\subsection{Synthetic datasets}
To evaluate our framework, we employed the synthetic networks derived from the four \textit{benchmark datasets} proposed in \cite{greene2010tracking}. These synthetic networks have been constructed through a dynamic extension of the static LFR benchmark~\cite{lancichinetti2009benchmarks} to model different types of evolutionary behaviors of communities over time. Each dataset, consisting of 15,000 vertices, contains five static networks (snapshots), simulating evolving communities. For each snapshot, the ground truth about static communities is available.
In each of the four synthetic networks, nodes have an average degree of $20$, a maximum degree of $40$, and a mixing parameter, $\mu=0.2$, that controls community overlap.
Furthermore, at each snapshot, $20\%$ of the nodes change their memberships, mirroring the natural migration of members between communities over time. The synthetic networks were designed to encompass all types of community evolution events:
\begin{itemize}
    \item \textit{BirthDeath}: 40 new communities are constructed to replace 40 existing communities.
    \item \textit{ExpandContract}: 40 randomly selected communities expand or contract their size by $25\%$ at each snapshot.
    \item \textit{MergeSplit}: 40 communities are randomly selected to be split, while another set of 40 undergoes merging, combining two communities into one.
    \item \textit{Intermittent}: at each snapshot, $10\%$ of the communities are intermittently concealed, making them unobservable.
\end{itemize}

\subsection{Evaluation metric and experimental setup}
To achieve a fair evaluation, we utilized the Louvain algorithm~\cite{blondel2008} as the static community detection approach for all the methods. We selected the hierarchical level that best matched the number of ground-truth static communities at the initial snapshot. To evaluate the performance in dynamic community tracking, we adopted an approach partially inspired by~\cite{greene2010tracking, HE2017438, folino2014}. For each evaluated competitor, dynamic communities were found considering two scenarios: i) using the ground-truth static communities at each snapshot; ii) using the static communities derived by Louvain as mentioned above. Then, we employed Normalized Mutual Information (NMI) to compare the memberships to dynamic communities found in both scenarios. 

NMI is a common entropy measure in information theory that quantifies the similarity between two clusters. It is defined as:
\begin{equation}
\begin{gathered}
N\!M\!I(X,Y) = \frac{H(X)+H(Y)-H(\!X,Y\!)}{(H(X)+H(Y))/2}
\end{gathered}
\end{equation}
where $H(X)$ is the entropy of the random variable $X$ associated with an identified community, $H(Y)$ is the entropy of the random variable $Y$ associated with a ground truth, and $H(X,Y)$ is the joint entropy.
A higher NMI value, nearing 1, indicates that the dynamic community detection is more resilient against noise in the static community information at each snapshot.
This resilience is crucial for ensuring reliable dynamic community tracking in real-world datasets.
 
For each dataset, we calculated the NMI in five sequential experiments, one per available snapshot. The first NMI value was found considering only the first snapshot, then for subsequent experiments we added one snapshot at a time. Hence, the last experiment employed the entire ``timeline'' made of five snapshots. 
As mentioned above, we compared our framework with other state-of-the-art approaches using the same Louvain-derived static community sets for unbiased comparison. Since some of these approaches require a threshold to be set, we tested them with different threshold values: (0.1; 0.3; 0.5). 
To select a threshold value, we calculated the average of the five NMI values for each dataset. Then, for each approach, we selected the threshold yielding the highest average NMI.

\subsection{State-of-the-art approaches}
\label{sec:algo}

In this section, we describe the state-of-the-art approaches that we compared against our framework: 
\textit{Greene}~\cite{greene2010tracking}, \textit{Takaffoli}~\cite{takaffoli2011, takaffoli2010framework}, \textit{Brodka (GED)}~\cite{brodka2013ged}, \textit{Tajeuna}~\cite{tajeuna2015tracking, tajeuna2016tracking}, and \textit{Mohammadmosaferi (ICEM)}~\cite{KADKHODAMOHAMMADMOSAFERI2020113221}. Similar to our framework, these methods begin by analyzing the static communities detected at each snapshot using a community detection algorithm.
Each of these approaches employs different strategies and similarity measures to track the evolution of communities over time, i.e., across snapshots. 

\textit{Greene et al.}~\cite{greene2010tracking} proposed a heuristic threshold-based method that enables many-to-many mappings between communities across different snapshots. The strategy proceeds as follows. First, an algorithm to find static communities is applied at each snapshot. Each community from the first snapshot is assigned to a dedicated dynamic community. Subsequently, communities at each snapshot are compared with each dynamic network's front community. The front community of a dynamic network is the community found at the most recent snapshot belonging to that dynamic community. To perform matching, the authors used the Jaccard coefficient for binary sets~\cite{10.2307/2427226}:
\begin{equation}
sim(\alpha, \beta) = \frac{|\alpha \cap \beta|}{|\alpha \cup \beta|}~.
\end{equation}
A pair is considered a match if its similarity surpasses a threshold $k$. Additionally, it is assumed that a community appearing at snapshot $t$ is deemed dissolved if it lacks a match for $d$ consecutive snapshots. This condition allows the discovery of evolving communities at non-consecutive snapshots. In our evaluation of this framework, we assumed $d=inf$ to enable matching with all non-consecutive snapshots. According to our experiments, the best result is obtained with $k=0.1$.

\begin{table}[!h]
\centering
\begin{tabular}{lrrr}
        \toprule
        \textbf{}        & \multicolumn{3}{c}{\textbf{Threshold}}  \\ 
        \cline{2-4}
        \textbf{Dataset} & \multicolumn{1}{c}{\textbf{0.1}} & \multicolumn{1}{c}{\textbf{0.3}} & \multicolumn{1}{c}{\textbf{0.5}} \\ 
        \midrule
        BirthDeath       & 0.984                  & 0.948                  & 0.926                  \\ 
        ExpandContract   & 0.986                  & 0.940                  & 0.921                  \\ 
        MergeSplit       & 0.969                  & 0.949                  & 0.936                  \\ 
        Intermittent     & 0.981                  & 0.936                  & 0.913                  \\ 
        \bottomrule
        \end{tabular}
        \caption{NMI values obtained by the Greene framework.}\label{tab:greene}
\end{table}

\textit{Takaffoli et al.}~\cite{takaffoli2011, takaffoli2010framework} used the following similarity measure to identify critical events in both consecutive and non-consecutive snapshots:
\begin{equation}
sim(\alpha, \beta) = \left \{ \begin{array}{ll}
\frac{|\alpha \, \cap \, \beta|}{max(|\alpha|, \, |\beta|)} \quad if \quad \frac{|\alpha \, \cap \, \beta|}{max(|\alpha|, \, |\beta|)} \geq k \\\\
0 \qquad \qquad \qquad \quad otherwise
\end{array}.
\right .
\end{equation}
The similarity threshold $k$ was automatically determined using a text-mining approach since it was assessed on networks with content information. Given our use of synthetic networks, we could not determine the threshold automatically. From our experiments, we obtained the best results using $k=0.3$ for this method.

\begin{table}[!h]
\centering
\begin{tabular}{lrrr}
    \toprule
    \textbf{}        & \multicolumn{3}{c}{\textbf{Threshold}} \\ 
    \cline{2-4}
    \textbf{Dataset} & \multicolumn{1}{c}{\textbf{0.1}} & \multicolumn{1}{c}{\textbf{0.3}} & \multicolumn{1}{c}{\textbf{0.5}} \\ \midrule
    BirthDeath       & 0.941                  & 0.950                  & 0.921                  \\ 
    ExpandContract   & 0.943                  & 0.946                  & 0.913                  \\ 
    MergeSplit       & 0.921                  & 0.928                  & 0.909                  \\ 
    Intermittent     & 0.940                  & 0.941                  & 0.903                  \\ 
    \bottomrule
    \end{tabular}
    \caption{NMI values obtained by the Takaffoli framework.}
    \label{tab:takaffoli}
\end{table}

\textit{Brodka et al.}~\cite{brodka2013ged} introduced the group evolution discovery (\textit{GED}) framework, designed to identify overlapping communities. 
Unlike previous methods that solely relied on a similarity metric to track community changes, this approach extends the measure by incorporating a topological metric:
\begin{equation}
I(\alpha, \beta) = \frac{|\alpha \, \cap \, \beta|}{|\alpha|} \cdot \frac{\sum_{x \in (\alpha \cap \beta)} NI_{\alpha(x)}}{\sum_{x \in (\alpha)} NI_{\alpha(x)}}
\end{equation}
where $NI_{\alpha}(x)$ represents the importance of node $x$ within the community $\alpha$. 
This value can be any centrality metric, e.g., degree centrality, social position, betweenness centrality, or PageRank.
Although the comparison is made at consecutive snapshots, the inclusion effect helps to track overlapping and non-overlapping communities. The method requires two threshold values, $k$ and $j$. We used the same value for $k$ and $j$ and obtained the best results for $k,j=0.1$.

\begin{table}[!h]
\centering
\begin{tabular}{lrrr}
        \toprule
        \textbf{}        & \multicolumn{3}{c}{\textbf{Threshold}} \\ 
        \cline{2-4}
        \textbf{Dataset} & \multicolumn{1}{c}{\textbf{0.1}} & \multicolumn{1}{c}{\textbf{0.3}} & \multicolumn{1}{c}{\textbf{0.5}} \\ \midrule
        BirthDeath       & 0.933                  & 0.931                  & 0.923                  \\ 
        ExpandContract   & 0.936                  & 0.932                  & 0.925                  \\ 
        MergeSplit       & 0.925                  & 0.923                  & 0.917                  \\ 
        Intermittent     & 0.926                  & 0.923                  & 0.918                  \\ 
        \bottomrule
        \end{tabular}
        \caption{NMI values obtained by the GED framework (Brodka et al.).}\label{tab:brodka}
\end{table}

\textit{Tajeuna et al.~\cite{tajeuna2015tracking, tajeuna2016tracking}} characterized each community using a \textit{transition probability vector}. These vectors encapsulate the extent of shared membership between different communities across time. To assess similarity between communities, the method compares their respective transition probability vectors.
Given two communities $\alpha$ and $\beta$ and their transition probability vectors $v_{\alpha}$ and $v_{\beta}$, the similarity between the communities is calculated as:
\begin{equation}
sim(\alpha, \beta) = \left \{ \begin{array}{ll}
\sum^{N_{c}}_{x=1} 2 \,
\frac{p_{\alpha, x} \, \cdot \, p_{\beta, x} }{p_{\alpha, x} \, + \, p_{\beta, x}} \\\\ \quad if \,\, \sum^{N_{c}}_{x=1} 2 \,
\frac{p_{\alpha, x} \, \cdot \, p_{\beta, x} }{p_{\alpha, x} \, + \, p_{\beta, x}} > k
\\\\
0 \qquad \qquad \qquad \quad otherwise
\end{array}
\right.
\end{equation}
where $p_{\alpha, x}$ and $p_{\beta, x}$ are components of the transition probability vectors $v_{\alpha}$ and $v_{\beta}$, respectively, and $N_{c}$ is the total number of communities in the dynamic network.
The threshold value $k$ is determined automatically as the approximated point of intersection of two Gamma curves. These curves derive from the non-zero similarity values between pairs of transition probability vectors.

\begin{table}[!h]
\centering
\begin{tabular}{lrll}
        \toprule
        \textbf{}        & \multicolumn{3}{c}{\textbf{Threshold}} \\ 
        \cline{2-4}
        \textbf{Dataset} & \multicolumn{3}{c}{Automatically set}                \\ \midrule
        BirthDeath       & \multicolumn{3}{r}{0.965}             \\ 
        ExpandContract   & \multicolumn{3}{r}{0.968}             \\ 
        MergeSplit       & \multicolumn{3}{r}{0.929}             \\ 
        Intermittent     & \multicolumn{3}{r}{0.965}             \\ 
        \bottomrule
        \end{tabular}
        \caption{NMI values obtained by the Tajeuna framework.}\label{tab:tajeuna}
\end{table}

\textit{Mohammadmosaferi et al.~\cite{KADKHODAMOHAMMADMOSAFERI2020113221}} introduced a novel method for Identification of Community Evolution by Mapping (\textit{ICEM}). In this method, the members of each community are mapped using a hash map. This map associates members with a pair consisting of the snapshot and a community index. From the second snapshot, ICEM builds a similarity list for each community and determines the evolution of a community based on that list. 
Besides identifying common critical events, the authors introduced the concept of \textit{partial} events. These events are identified based on the partial similarity between two communities:
\begin{equation}
sim(\alpha, \beta) = \frac{|\alpha \cap \beta|}{|\alpha|}
\end{equation}
\begin{equation}
sim(\beta, \alpha) = \frac{|\alpha \cap \beta|}{|\beta|}
\end{equation}
where $\alpha$ is a community from the $i^{th}$ snapshot and $\beta$ a community from the $j^{th}$ snapshot, where $i < j$.
The communities $\alpha$ and $\beta$ are partially similar if $sim(\alpha, \beta) > k$ and $sim(\beta, \alpha) > k$, while they are very similar if $sim(\alpha, \beta) > v$. Therefore, $k$ and $v$ are the  thresholds to identify partially similar and very similar communities, respectively. Since our concern is not to distinguish different types of critical events in this evaluation, we set $v=0.5$ and obtained the best results for $k=0.1$.

\begin{table}[!h]
\centering

\begin{tabular}{lrrr}
\toprule
\textbf{}        & \multicolumn{3}{c}{\textbf{Threshold}} \\ 
\cline{2-4}

\textbf{Dataset} & \multicolumn{1}{c}{\textbf{0.1}} & \multicolumn{1}{c}{\textbf{0.3}} & \multicolumn{1}{c}{\textbf{0.5}} \\ \midrule
BirthDeath      & 0.973 & 0.972 & 0.936 \\ 
ExpandContract  & 0.975 & 0.970 & 0.930 \\ 
MergeSplit      & 0.935 & 0.956 & 0.923 \\ 
Intermittent    & 0.974 & 0.967 & 0.922 \\ 
\bottomrule
\end{tabular}
\caption{NMI values obtained by the ICEM framework (Mohammadmosaferi et al.).}
\label{tab:mohammadmosaferi}
\end{table}

\begin{table}[!h]
\centering

\small
\begin{tabular}{lrrrrr}
\toprule
                 & \multicolumn{5}{c}{\textbf{Snapshot}} \\ 
                 \cline{2-6}
\textbf{Dataset} & \multicolumn{1}{c}{1} & \multicolumn{1}{c}{2} & \multicolumn{1}{c}{3} & \multicolumn{1}{c}{4} & \multicolumn{1}{c}{5} \\ \midrule
  BirthDeath      & 0.971 & 0.968 & 0.985 & 0.985 & 0.979 \\ 
  ExpandContract  & 0.971 & 0.993 & 0.995 & 0.993 & 0.987 \\ 
  MergeSplit      & 0.971 & 0.988 & 0.977 & 0.979 & 0.966 \\ 
  Intermittent    & 0.971 & 0.973 & 0.991 & 0.991 & 0.979 \\ 
  \bottomrule
\end{tabular}
\caption{NMI values obtained by our framework.}
\label{tab:results}
\end{table}

\begin{figure}[t]
  \centering

  \subfloat[Birthdeath]{
  \includegraphics[width=0.5\columnwidth]{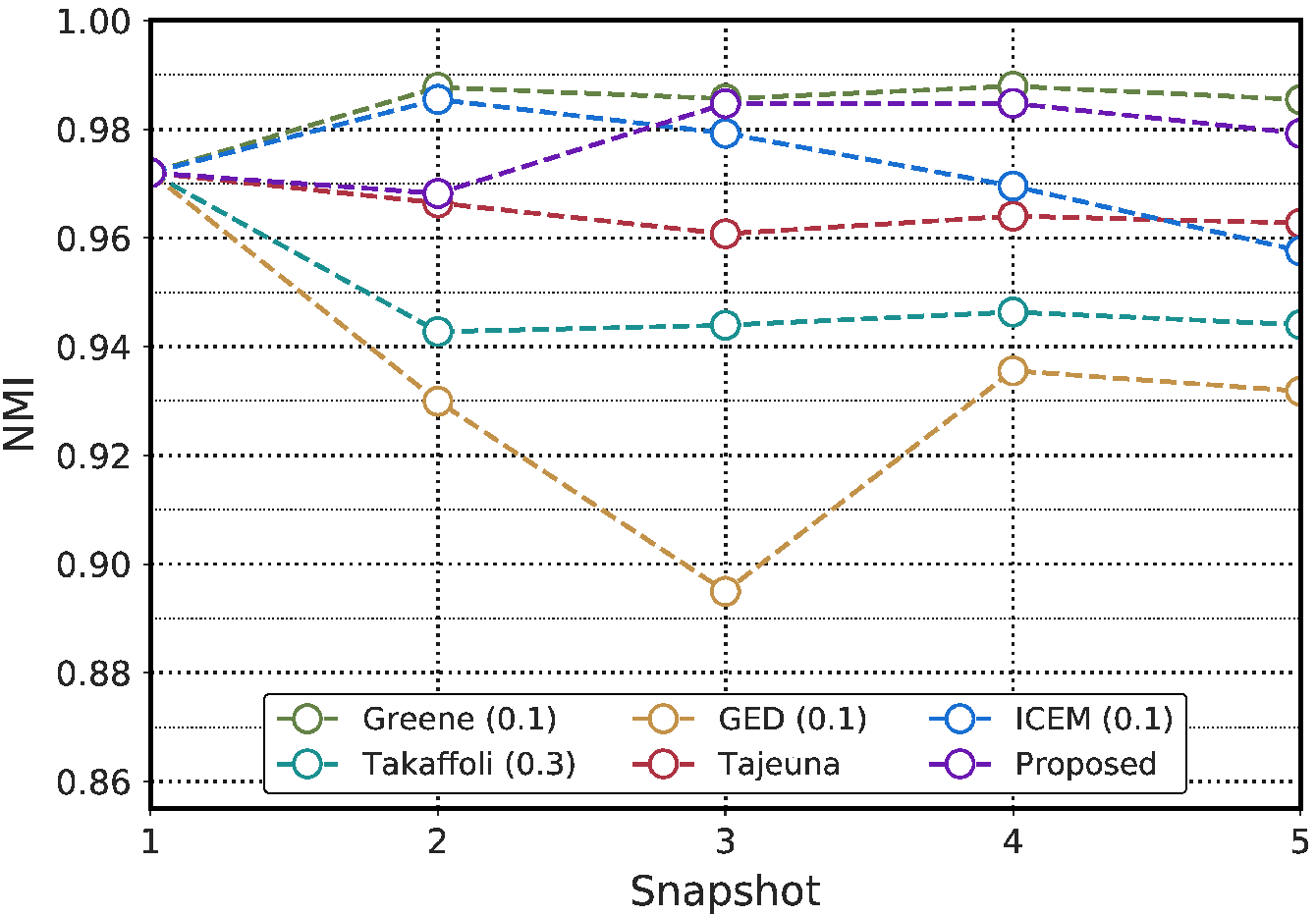}
  }
  \subfloat[ExpandContract]{
  \includegraphics[width=0.5\columnwidth]{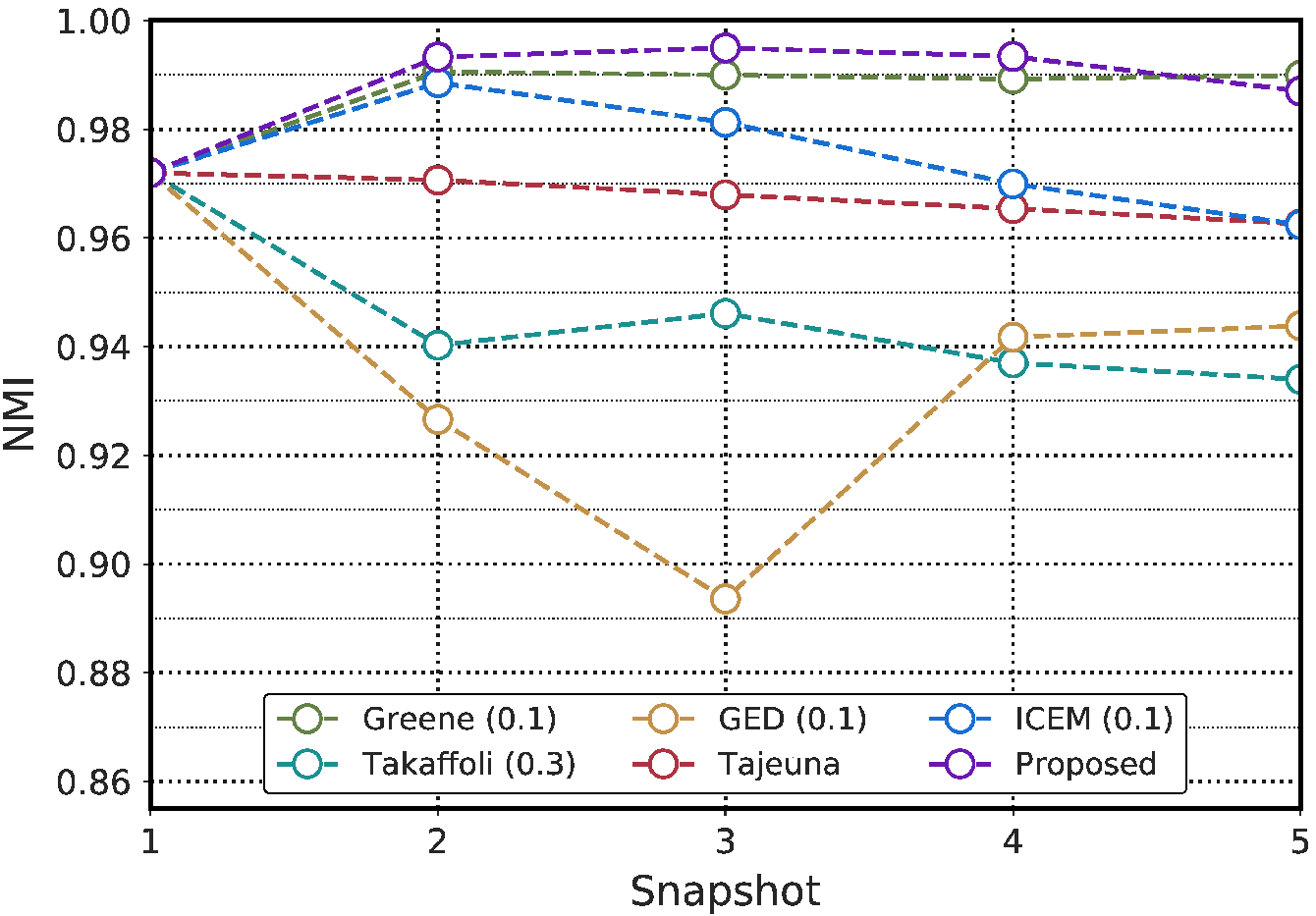}
  }
  \hfill
  \subfloat[MergeSplit]{
  \includegraphics[width=0.5\columnwidth]{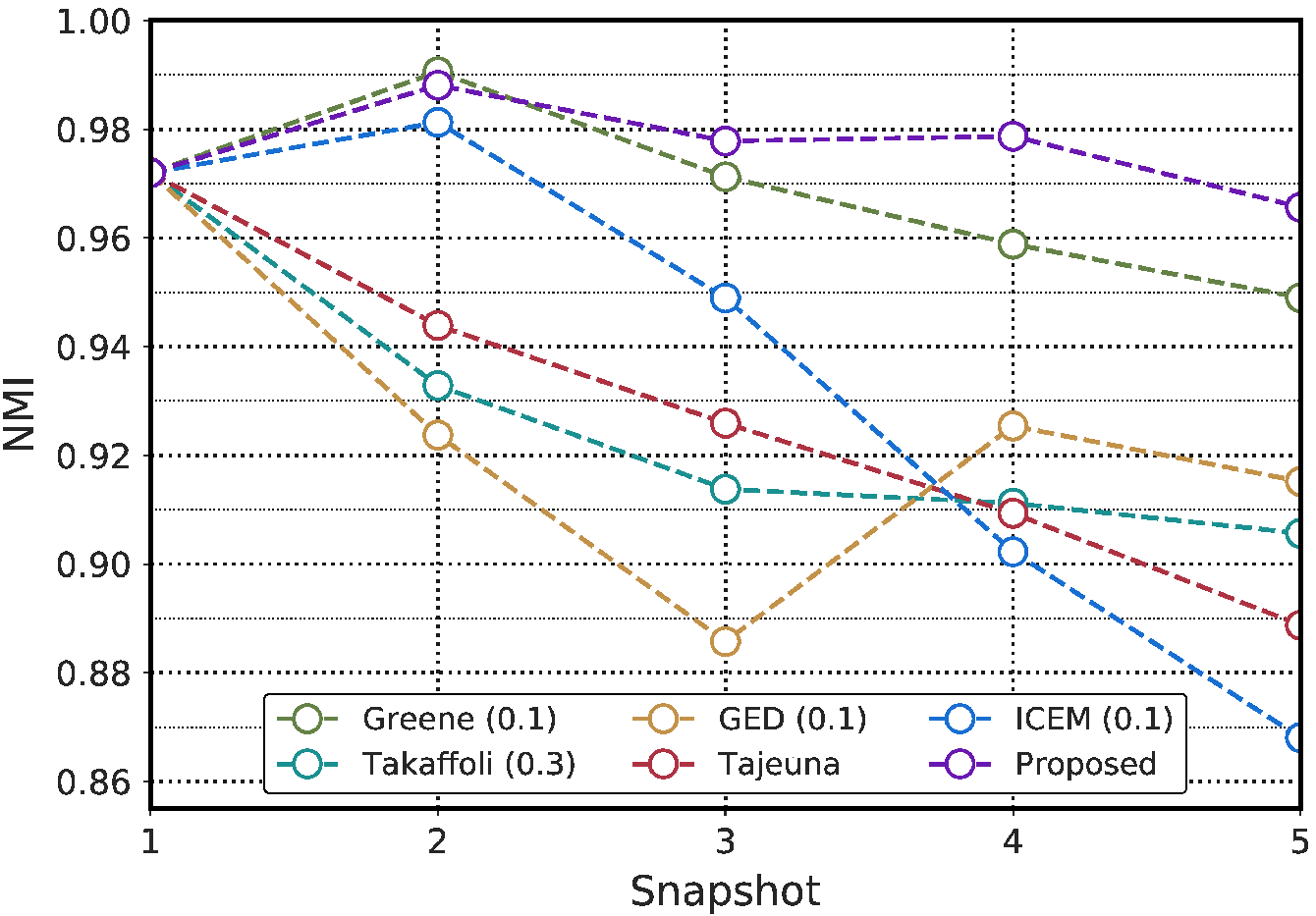}
  }
  \subfloat[Intermittent]{
  \includegraphics[width=0.5\columnwidth]{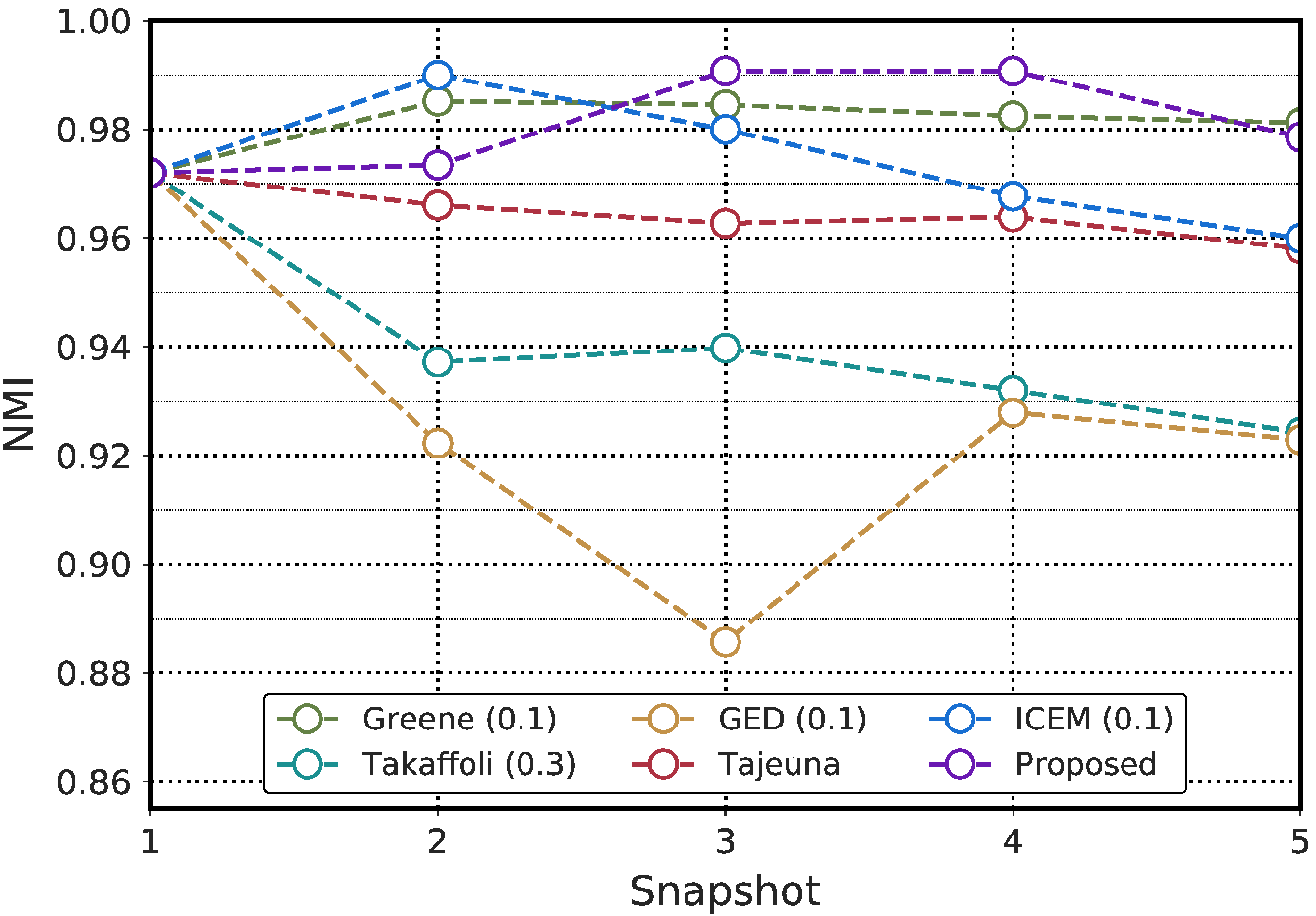}
  }
  
  \caption{NMI of the proposed dynamic community tracking framework and state-of-the-art methods on the four synthetic datasets.}
  \label{fig:results}
\end{figure}

\subsection{Results of the evaluation and comparison}

Table~\ref{tab:results} displays the NMI values achieved by our framework across all datasets and snapshots. Notably, it secured high NMI values (above 0.96) in every experiment.

Figure~\ref{fig:results} contrasts our results with the state-of-the-art methods on the four synthetic datasets. Our approach offers the best performance in most scenarios, although Greene and ICEM occasionally achieve a slightly higher NMI value. Overall, Greene, ICEM, and Tajeuna, exhibit strong results, whereas Takaffoli and particularly GED fall short in most experiments. GED's underwhelming performance may stem from its inability to link non-consecutive communities.
Interestingly, all the evaluated methods achieve their worst performance with the \textit{MergeSplit} dataset. Merge and split events significantly distort the dynamic community structures, making them challenging to track. Nevertheless, our method consistently achieves an NMI score above 0.96 even for this dataset. Among others, only Greene managed to retain NMI values above 0.94 in this dataset, while ICEM and Tajeuna show a degrading performance as the number of available snapshots increases.
For scenarios with less complex evolutionary patterns, the efficacy of our method becomes more pronounced with the inclusion of more than two snapshots.
A similar trend is also observed with Greene. With the expansion of the community similarity network, pinpointing analogous community clusters with higher precision becomes more feasible. In contrast, some competing methods exhibit increased vulnerability to community matching errors as the snapshot count grows.

\begin{figure}[t]
  \centering
  \includegraphics[width=0.5\linewidth]{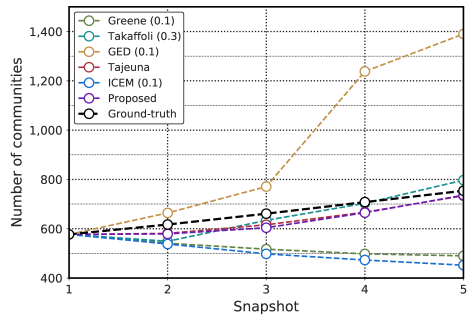}
  \caption{Number of dynamic communities found on the \textit{MergeSplit} synthetic dynamic network, by the proposed dynamic community tracking method and state-of-the-art methods, relative to the ground-truth number of dynamic communities.}
  \label{fig:n_comm}
\end{figure}

Figure~\ref{fig:n_comm} offers a distinct perspective on the performance of the evaluated methods within the \textit{MergeSplit} dataset. It contrasts the count of ground-truth dynamic communities in each snapshot with those identified by the various methods. Our method, together with Tajeuna and Takaffoli, closely mirrors the ground truth in terms of the number of dynamic communities. The discrepancy observed with Greene and ICEM could be attributed to the low threshold value employed, which results in increased community matches and subsequent merging. In contrast, GED, as previously mentioned, suffers from its limited comparison scope, restricted to consecutive snapshot communities.

In summary, the proposed framework excels in handling dynamic community tracking by focusing on local modularity optimization and setting aside the temporal component. This unique design choice allows the framework to adapt to structural changes in dynamic communities, consistently delivering strong results. Unlike most competitors, our approach operates without the need to set a threshold value. Although this entails the drawback that an unsatisfactory result cannot be altered by varying the threshold value, it is crucial to acknowledge that community tracking is a complex process with no universally correct outcomes. As such, the absence of a threshold can be seen as an advantage, preventing misleading adjustments. While methods that permit threshold adjustments might yield seemingly satisfactory results, they may not truly reflect the genuine evolution of dynamic communities. Additionally, by leveraging a denser community similarity network, our framework achieves enhanced performance when using more than two snapshots.

\subsection{Complexity analysis}

We excluded the computational cost of network representation and static community detection from our complexity analysis, as our framework builds upon communities derived from any given static community detection method. In our framework, each static community is compared with all others in subsequent snapshots, yielding a time complexity of $O(c^2)$, with $c$ representing the total number of static communities. The local modularity optimization phase takes $O(c)$. Thus, the total time and memory complexity of the framework is $O(c^2)$. To assess the scalability of our framework, we employed the synthetic graph generator proposed by~\cite{greene2010tracking} to create progressively larger dynamic networks. Table~\ref{tab:time} reports the properties of the datasets along with the corresponding execution times. We generated synthetic dynamic networks in two sets: one with five snapshots and another with ten. The number of nodes ranges from 1,000 to 100,000, which determines the number of static communities handled by the framework. We used a Python implementation of the proposed framework and executed the experiments using a single core of a machine with an Intel Xeon Processor (2.2~GHz) and 64~GB of RAM. For a dynamic network with 100,000 nodes across five snapshots, the processing time is under 35 minutes. However, with ten snapshots, it increases to approximately two and a half hours. The execution time is heavily influenced by the number of snapshots in the dynamic network, given its positive correlation with the count of static communities. Since most computations, such as determining the similarity between two communities, can operate independently, there is potential to boost computational performance through parallelization. Further improvements could be achieved by combining Minhashing~\cite{broder1997on} and Local Sensitive Hashing (LSH)~\cite{piotr1998approx} to identify approximate nearest neighbors in the Jaccard space. This would significantly reduce the number of comparisons, edging closer to linear time complexity.

\begin{table}[ht]
\centering
\begin{tabular}{rrrr}
\toprule
\multicolumn{3}{c}{\textbf{Dataset Characteristics}}  & \\
\cline{1-3}
\multicolumn{1}{r}{snapshots}  & \multicolumn{1}{r}{nodes}  & communities & \multicolumn{1}{c}{\textbf{Time}} \\ 
\midrule
 
\multicolumn{1}{c}{\multirow{3}{*}{5}} & \multicolumn{1}{r}{1,000} & 438 & \multicolumn{1}{r}{$<$ 1 s}  \\ 

\multicolumn{1}{c}{} & \multicolumn{1}{r}{10,000} & 4,369    & \multicolumn{1}{r}{21 s}  \\ 

\multicolumn{1}{c}{}            & \multicolumn{1}{r}{100,000}    & 43,704   & \multicolumn{1}{r}{34 m 45 s} \\ 
\midrule

\multicolumn{1}{c}{\multirow{3}{*}{10}}   & \multicolumn{1}{r}{1,000}      & 872      & \multicolumn{1}{r}{1 s} \\ 

\multicolumn{1}{c}{}           & \multicolumn{1}{r}{10,000}     & 8,738    & \multicolumn{1}{r}{1 m 33 s} \\ 

\multicolumn{1}{c}{}           & \multicolumn{1}{r}{100,000}    & 87,559   & \multicolumn{1}{r}{2 h 31 m 53 s} \\
\bottomrule
\end{tabular}
\caption{Execution time of the proposed framework on networks with different characteristics.}
\label{tab:time}
\end{table}

\subsection{Application to a real-world dataset}

For our second evaluation, we assessed the behavior of the proposed framework using a real dataset described by~\cite{mazza2022}. This dataset comprises the activity of 63,358 fake Twitter accounts, which produced 5,457,758 tweets during 2020. The term ``fake account'' refers to social media accounts that contain false information or pretend to be a real person or organization~\cite{mazza2_2022}. This dataset is ideal for our experiments because the tweets were not confined to specific languages or topics, providing a broad range of content that included, for instance, politics and cryptocurrency~\cite{colaITASEC23}. We set snapshot granularity to one day of activity, and produced a weighted undirected graph for each snapshot. Daily graphs were derived from a co-hashtag network, in which two users are linked if they both used the same hashtag on a given day. The weight of each edge is given by the number of hashtags used in common between two users. To find the static communities at each snapshot, we used the Order Statistics Local Optimization Method (OSLOM)~\cite{lancichinetti2011}, which is based on measuring the  significance of communities compared to a null model without community structure~\cite{molloy1995}. After OSLOM identifies statistically significant communities by grouping neighboring nodes, it undergoes multiple iterations of adjustments, like node addition or deletion, to enhance the significance of communities. We selected OSLOM due to its noted efficacy in detecting communities within online social networks~\cite{darmon2015, wendel2017}. A possible drawback of OSLOM is its tendency to identify smaller static communities~\cite{KUMAR2017103}, which may not properly capture the community structure of a network~\cite{SHI2013394}. This concern is not relevant to the experiment, as the identified static communities are joined into bigger dynamic communities by the proposed framework.

We applied our framework to the static communities identified by OSLOM, obtaining 103 dynamic communities.
To investigate the temporal behavior of the identified dynamic communities, we identified the hashtags used daily within each static community.
Then, we introduced a new metric related to a dynamic community: the \textit{average hashtags overlap}. Given two sets of hashtags $h_1$ and $h_2$, the hashtags overlap is found as:
\begin{equation} 
Hashtags \, Overlap = \frac{|h_1 \, \cap \, h_2|}{min(|h_1|, |h_2|)}~.
\end{equation}
The average hashtags overlap for a dynamic community $D$ is calculated by averaging the hashtags overlaps across all its constituent static communities.
An acceptable alternative metric might have been the one used in~\cite{takaffoli2011}, where instead of keywords or hashtags, the authors relied on topics. Given that this dataset captures the activity of sold fake accounts over a year, it is plausible that communities employed varied hashtag sets on different days. Despite this, a set of hashtags effectively represents a community on a specific day. 

\begin{figure}[ht]
    \centering
  \includegraphics[width=0.5\linewidth]{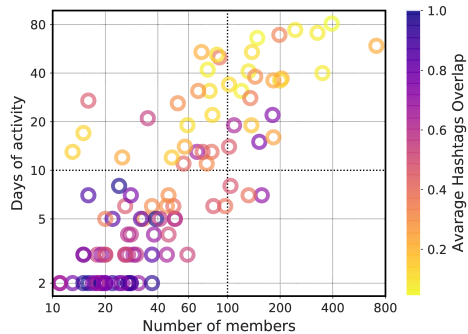}
  \caption{Circles represent dynamic communities, plotted with the overall number of members on the $x$ axis against the days of activity on the $y$ axis. The color of each circle encodes its average hashtags overlap value.}
  \label{fig:mut_top}
\end{figure}

Averaging the hashtags overlap values for all static communities within a dynamic community allows us to gain insights into the characteristics of the identified dynamic communities.
Figure~\ref{fig:mut_top} plots the identified dynamic communities, with the overall number of members on the $x$ axis and the days of activity on the $y$ axis. Moreover, through color, we reported the average hashtags overlap of each dynamic community. 
This plot reveals a correlation between the number of members and days of activity; specifically, communities with more users tend to be active for more days, and vice versa.
In addition, dynamic communities with high values in terms of members and days of activity also exhibit a low value of average hashtags overlap, meaning that during the one-year period, these dynamic communities discussed disparate topics.  A low value of average hashtags overlap does not imply that a dynamic community is not relevant, as this metric may be affected by other factors, such as the strategy employed.

\begin{figure*}[ht]
  \includegraphics[width=\textwidth]{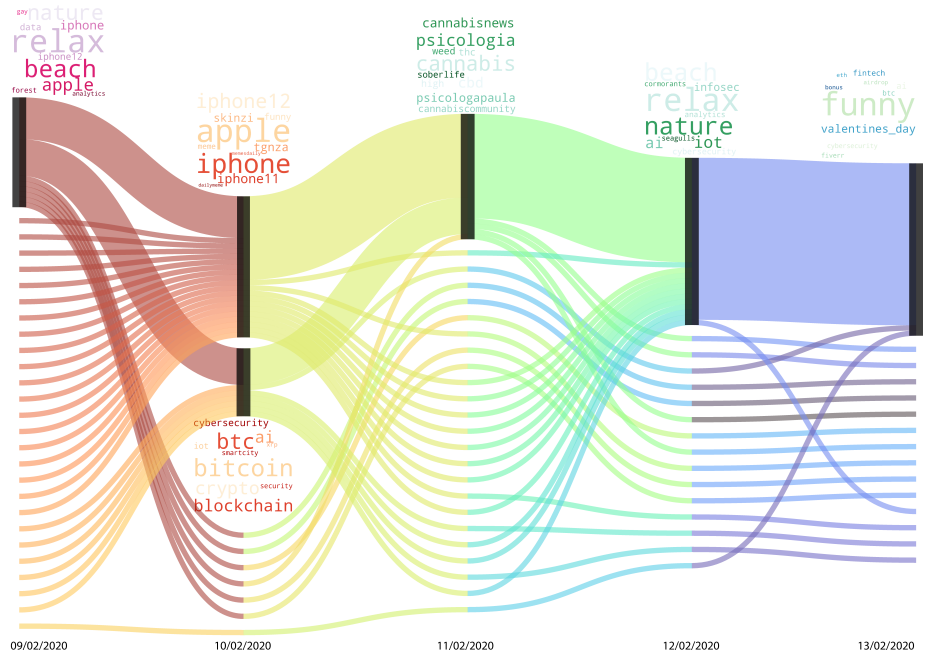}
  \caption{Alluvial diagram representing the structure of the analyzed dynamic community.}
  \label{fig:alluvial}
\end{figure*}

Figure~\ref{fig:alluvial} provides a glimpse into a dynamic community's activity from February 9 to February 13, 2020. To better understand temporal behavior, for each constituent community we show the ten most used hashtags. We observe that initially, there is only one community using mostly generic hashtags such as \#relax, \#beach, and \#nature and some more specific ones such as \#apple and \#iphone. The next day, this community split into two communities, one of which turns out to be larger than the initial one due to the arrival of new users. This community resumes the hashtags used in the previous day regarding Apple products, introducing new ones such as \#iphone11 and \#iphone12. Instead, the other community focuses on cryptocurrency and blockchain technology through the hashtags \#btc, \#bitcoin, and \#blockchain. On the following day, a portion of users from these two communities merge into a new community centered on hemp-related topics: \#cannabis, \#cbd, \#cannabisnews. However, this community expands on the subsequent day, reverting to more generic hashtags. This behavior continues the next day, where we find \#funny as well as \#valentines\_day since it is Valentine's Eve.

This example highlights the variability in topics that users of an online social network can address from one day to the next. Observing this kind of phenomenon and, more generally, studying the behavior of a community over time is only possible by taking into account its temporal properties, that is, by analyzing it as a dynamic community. 
By extending the framework's application to similar datasets~\cite{cola2023twitter, chen2023tweets}, analysts could gain new time-related insights into various phenomena on social media platforms, such as coordinated inauthentic behaviors.

\section{Conclusions}
\label{sec:conclusions}

In this paper, we introduced a framework designed to track the evolution of dynamic communities over time. Unlike most existing frameworks, our community matching phase operates without a threshold value. Instead, it employs modularity optimization applied to a weighted, undirected community similarity network, where the temporal component is omitted. 
We assessed our framework using synthetic graphs with embedded events and benchmarked our results against other state-of-the-art frameworks. Our method excelled in handling structural changes in dynamic communities, yielding consistently strong results across diverse scenarios. The absence of a threshold further ensures the consistency of these results.
Moreover, our framework is agnostic to the specific community detection algorithm used, offering flexibility in algorithm choice based on network properties, such as whether the network is weighted or unweighted, directed or undirected. 
We also conducted a preliminary evaluation of the proposed framework on a real-world Twitter network, uncovering 103 dynamic communities with distinct characteristics. This finding suggests that dynamic communities, particularly those derived from online social media, can show considerable variability in the topics they encompass.
When applied to online social media analysis, our framework shows promising potential in identifying user groups with aligned interests or behaviors. By tracking dynamic communities, our approach can also enhance the temporal analysis of influencers and key opinion leaders, offering valuable insights for research applications.

\section*{Acknowledgements}

This work was supported by: the European Union – NextGenerationEU – National Recovery and Resilience Plan (Piano Nazionale di Ripresa e Resilienza, PNRR) – Project: “SoBigData.it – Strengthening the Italian RI for Social Mining and Big Data Analytics” – Prot. IR0000013 – Avviso n. 3264 del 28/12/2021; project SERICS (PE00000014) under the NRRP MUR program funded by the EU - NGEU.

\linespread{1.1}\selectfont
\bibliographystyle{elsarticle-num}
\bibliography{article}

\end{document}